\documentclass[useAMS,usenatbib]{mn2e}
\usepackage{epsfig}

\def\equ#1{eq.~(\ref{eq:#1})}
\def\se#1{\S\ref{sec:#1}}
\def\Fig#1{Fig.~\ref{fig:#1}}
\def\etal{{\it et al.\ }}

\def\be{\begin{equation}}
\def\ee{\end{equation}}
\def\ba{\begin{eqnarray}}
\def\ea{\end{eqnarray}}
\def\prop{\propto}
\def\ifm#1{\relax\ifmmode#1\else$\mathsurround=0pt #1$\fi}
\def\kms{\ifmmode\,{\rm km}\,{\rm s}^{-1}\else km$\,$s$^{-1}$\fi}

\def\kpc{\,{\rm kpc}}

\def\ltsima{$\; \buildrel < \over \sim \;$}
\def\lsim{\lower.5ex\hbox{\ltsima}}
\def\gtsima{$\; \buildrel > \over \sim \;$}
\def\gsim{\lower.5ex\hbox{\gtsima}}

\def\omm{\Omega_{\rm m}}
\def\omb{\Omega_{\rm b}}
\def\oml{\Omega_{\Lambda}}
\def\om0{\Omega_{{\rm m} 0}}
\def\ol0{\Omega_{\Lambda 0}}

\def\Vv{V_{\rm v}}
\def\Mv{M_{\rm v}}
\def\Rv{R_{\rm v}}

\def\Ms{M_{\rm *}}

\def\rc{r_{\rm c}}
\def\Mc{M_{\rm c}}

\def\zion{z_{\rm ion}}
\def\zend{z_{\rm end}}
\def\Tion{T_{\rm ion}}
\def\Vev{V_{\rm evap}}

\title[Photo-evaporation in dwarf galaxies]
{Photo-evaporation by thermal winds in dwarf galaxies}

\author[N. J. Shaviv \& A. Dekel]
{Nir J. Shaviv \& Avishai Dekel \\
Racah Institute of Physics, The Hebrew University, Jerusalem 91904, 
Israel}

\begin{document}

\pagerange{\pageref{firstpage}--\pageref{lastpage}} \pubyear{2002}
\maketitle
\label{firstpage}

\begin{abstract}

We revisit the evaporation process of gas from dwarf galaxies after it
has been photo-ionized by the UV flux from the first stars and AGNs
and heated to $T \simeq 10^4$ or $2 \times 10^4$K respectively.
Earlier estimates, based on the balance between pressure and gravity,
indicated that dark haloes of virial velocity lower than
$\Vev \simeq 11-13\kms$ have lost most of their gas in a dynamical time.
We follow the continuous evaporation by a thermal wind during the 
period when the ionizing flux was effective.
We find that the critical virial velocity for significant evaporation is
significantly higher.
For example, if the ionization starts at $\zion=10$ and is maintained 
until $z=2$, a mass loss of one $e$-fold occurs in haloes of
$\Vev \simeq 25$ (or $35 \kms$) 
for $T \simeq 10^4$K (or $2 \times 10^4$K).
Haloes of $\Vev \simeq 21\kms$ (or $29\kms$)
lose one $e$-fold within the first Hubble time at $z=10$.
Any dwarf galaxies with virial velocities smaller than $\Vev$ must have formed
their stars from a small fraction of their gas before $\zion$, and then lost
the rest of the gas by photo-evaporation.
This may explain the gas-poor, low surface brightness dwarf spheroidal
galaxies.
By $z<1$, most of the IGM gas was evaporated at least once form
dwarf galaxies, thus providing a lower bound to its metallicity.
\end{abstract}


\section{INTRODUCTION}
\label{sec:intro}

The first stars and AGNs generate a flux of UV radiation which 
photoionizes
most of the hydrogen in the universe by $\zion \sim 7$
\citep[see a review by][]{barkana:01}.
The gas is heated to a temperature of $\Tion \simeq 10^4$K
and is kept at this temperature as long as the ionizing flux persists.
The actual temperature mostly depends on the hardness of the radiation
flux and therefore on whether its origin is in stars or AGNs, 
respectively, but also on the flux itself. 
The total emissivity of the observed quasars at $z=3$ is probably 
sufficient to have completely reionized the low density intergalactic 
Helium already before that redshift, thereby increasing the temperature 
to $\Tion \simeq 2 \times 10^4$K \citep{miralda:00,barkana:01}.
Dark-matter haloes with an escape velocity lower than the associated 
thermal velocity of the ionized gas, on the order of $10\kms$,
cannot retain this hot gas --- it would photo-evaporate.
\citet{barkana:99} have estimated the gas loss based on the balance
between pressure and gravity, and obtained that the haloes that lose 
50\% of their gas in a dynamical time are of a typical virial velocity of
$\Vev \simeq 11-13\kms$.

Based on this low estimate of the critical velocity for 
photo-evaporation,
one could have concluded that photo-evaporation is not
a very important process in galactic history. This is because a similar 
lower
bound on the virial velocity of haloes that can form stars is obtained
from simple cooling arguments. The radiative cooling rate,
which is very high just above $10^4$K due to atomic recombination,
becomes extremely low just below $10^4$K.
The only available cooling agents in this range are $H_2$ molecules,
which are both inefficient and fragile in the presence of even a weak 
UV flux
\citep{haiman:96}.
This implies that haloes of virial velocity below $\sim 10\kms$ cannot
have their gas cool to form stars independently of the radiative 
feedback
and the associated reionization.
Indeed, there is an indication for such a lower bound in
the population of Local Group dwarf galaxies \citep[see][]{woo:03}.

Another scale associated with the radiative feedback process is the 
Jeans
scale. Once the IGM is photo-ionized, its pressure shuts off any further
gas infall into haloes of virial velocities up to $\sim 30\kms$
\citep[see a review of numerical results in][\S 6.5]{barkana:01}.
Gas could resume falling into small halos after $z \sim 1-2$ when the UV
background flux declined sufficiently \citep{babul:92}, but only
halos of $V>20-25\kms$ can form molecular hydrogen by $z\sim 1$
\citep{kepner:97}.
This is therefore a lower bound for galaxies unless they managed to
have their gas fall into virialized haloes before $\zion$.

The observed internal velocities in Local-Group dwarf galaxies
indicate that dwarf galaxies with virial velocities in the range
$10-30\kms$ do exist.
However, a comparison of the galaxy luminosity function at the faint end
(with a logarithmic slope close to $-1$)
and the halo mass function predicted by the standard $\Lambda$CDM 
cosmological
model (with a log slope near $-2$)
indicates that some haloes in this range must have remained dark
with no stars in them.
Another relevant observation is that
most of the galaxies in this range are dwarf spheroidals, showing only 
little
gas and current star formation.
This calls for a more careful analysis of the fate of the ionized gas in
such haloes.
Since an effective ionizing flux may persist until $z \sim 1-2$
\citep{babul:92}, the evaporation process is likely to continue for
many dynamical times in the form of a continuous thermal wind,
and thus end up with more mass loss than estimated by 
\citet{barkana:99},
and correspondingly with a higher critical velocity $\Vev$
for significant evaporation.
This wind is analogous to the Parker wind from the sun, which is
responsible for a mass loss of $10^{-14} M_\odot/yr$ and is well 
understood
\citep[e.g.,][]{lamers:99}.
With a higher $\Vev$, the photo-evaporation would become a major player
in the formation of dwarf galaxies.
In haloes smaller than this critical velocity (but larger than $\sim 
10\kms$),
stars can form only before $\zion$, and any left-over gas cannot be 
retained.
This could lead to the formation of gas-poor dwarfs, not unlike the 
dwarf spheroidals
observed in the Local Group, or to completely dark haloes also in the 
range
$10<V<\Vev$.
If a significant fraction of the universal gas was able to evaporate 
from haloes
that managed to form stars earlier, then the evaporation process might 
also
have had implications on the metal enrichment of the IGM.
We therefore attempt to evaluate the critical virial velocity for 
evaporation
by pursuing a dynamical calculation of the evaporation process.

In \se{equations} we present the basic treatment of thermal winds
applicable to non-point like mass distributions.
In \se{nfw}
we apply the wind analysis to cosmological galactic haloes.
In \se{conc} we discuss our results.

\section{Wind Analysis}
\label{sec:equations}

We wish to compute the mass-loss rate in a steady isothermal
thermal
wind in a galactic halo.
The formulation is inspired by the standard problem of isothermal winds
from stellar objects \citep{parker:60,brandt:70,lamers:99}.
The difference is largely in the nomenclature.
The main mathematical modification that we have to consider is that
the haloes are not point like objects, that is, the gravitational force
per unit mass is now $-G M(r)/r^2 \hat{\bf r}$ with a monotonically
increasing $M(r)$.

We begin by considering an arbitrary spherical potential $\phi(r)$.
The gas is assumed to be heated and kept at a constant temperature $T$,
typically $\sim 10^4$K.
\citet{barkana:99} calculated the amount of gas that unbinds 
``instantaneously" as a result of the extra thermal energy. All the 
leftover gas,
which energetically cannot escape, was assumed to settle down in the
haloes' potential well.

\citet{parker:60} found that even if the gas is energetically bound,
a wind can develop. One way of looking at it is that irrespective of how
deep the potential well is, an exponentially small fraction of
the gas can escape.
Without continuous heating, a thermal wind would progressively cool as
it expands, the pressure pushing it out would decrease,
and the wind would die out.
On the other hand, if the gas can constantly be heated to compensate 
for the
adiabatic cooling, it is easier for the gas to be expelled, and a large 
flux
can be obtained and maintained.

We assume
in the following analysis that the wind is kept at a fixed temperature.
This follows from the assumption that over the time scale it takes the
gas to accelerate
out of the potential well, the radiative equilibrium with the ionizing
radiation can keep the gas warm.  We
will check the consistency of our results with this assumption
and discuss its meaning
in \se{isothermality}.

The basic equations are the continuity equation,
the Euler equation of motion, and the equation of state of an
isothermal gas:
\be
{\partial \rho \over \partial t} = - \nabla \cdot (\rho {\bf v})  ,
\label{eq:cont}
\ee
\be
\rho {D {\bf v} \over D t}  = - \nabla P + {\bf f}_{grav} ,
\label{eq:euler}
\ee
\be
P = c_s^2 \rho ,
\label{eq:state}
\ee
where $c_s$ is the isothermal speed of sound.
Under the assumption of spherical symmetry, the equations become
\be
\dot\rho = -\frac{1}{r^2} (r^2 \rho v)' ,
\label{eq:cont_sph}
\ee
\be
\dot v +v v' = -c_s^2 \frac{\rho'}{\rho} -\phi' ,
\label{eq:euler_sph}
\ee
where $\phi(r)$ is the gravitational potential normalized to 0 at $r=0$:
\be
\phi(r) = \int_{0}^{r} {G M(r) \over r^2} dr,
\ee
and where dot and prime denote Eulerian partial derivatives with 
respect to $t$
and $r$ respectively.

We now assume {\em steady state}, where $\dot M$ is the same at every 
$r$.
Writing $M=\int 4\pi r^2 \rho dr$, and using continuity, we obtain
$\dot M = -4\pi r^2 \rho v$. Steady state therefore implies $r^2 \rho v
=$const., and then from continuity $\dot \rho=0$. It also requires 
$\dot v=0$.
The equations then combine to the {\em wind equation}:
\be
\label{eq:wind1}
\left( v - {c_s^2 \over v}\right) v' = -\phi'(r) + {2 c_s^2 \over r} .
\ee

Any {\em steady state} theory with a {\em transonic} solution, such as a
thermal wind, necessarily requires that the transition between the 
sub-sonic
and super-sonic flow will either take place at a shock
(which is of no interest for us here)
or that the sonic point will coincide with the critical point
where the net radial forces vanish.
This is because a transonic solution to eq.~\ref{eq:wind1} implies that
the left-hand side changes its sign, and therefore the right-hand side
has to change sign at the same radius.\footnote{There are some 
exceptions
to this line of arguments, for example when the force is a function of
velocity.} Thus, the pressure force has to balance the gravitational 
force
at the sonic point.

The {\em sonic point} $r_s$ can therefore be obtained from the given 
potential
profile and the given speed of sound by
\be
\label{eq:sonic-def}
\phi'(r_s) = {2 c_s^2 \over r_s} .
\ee

Generally, the solution of this equation is expected to be implicit
and it should therefore be solved numerically.
Nevertheless, one can anticipate a solution of order
$r_s \sim {G M / c_s^2}$
(where $M$ is some typical mass within the sonic radius).
In other words, the escape velocity from the sonic radius will generally
be of the order of the speed of sound.

We next wish to obtain the {\em velocity profile} of the wind.
By integrating the Euler equation over $r$, with the steady-state 
condition
$\dot v=0$, we obtain
\be\label{ieq:sonic}
{1 \over 2} v^2(r) + {c_s^2 \ln \rho(r)} + \phi (r) = {\rm const.},
\ee
a constant in $r$.
By integrating the continuity equation, with the steady-state condition
$\dot \rho =0$, we obtain
\be\label{}
\ln \rho(r) + 2\ln r +\ln v(r)  = {\rm const.}
\ee
When combined, the two equations yield:
\be\label{}
{1\over 2}{v^2(r) \over c_s^2} - \ln v(r) - 2 \ln r +
{\phi (r) \over c_s^2} =  {\rm const.}
\ee
The constant is determined by the requirement that $v=c_s$ at the sonic 
point.
In exponential form, the equation becomes
\be
\label{eq:velocity}
{v(r)\over c_s} \exp \left(- {v^2(r) \over 2 c_s^2}\right)
= \left( r_s\over r\right)^{2} \exp \left({\phi (r) \over c_s^2}
-{\phi (r_s) \over c_s^2}- {1\over 2} \right) .
\ee
\vskip -2mm This equation relates the velocity profile and the potential
profile, where $r_s$ is obtained from \equ{sonic-def}.

The {\em mass loss rate} can now be evaluated by equating its value at 
the
sonic point and at some small intermediate radius $r_i$ where $v \ll 
c_s$,
using the wind velocity profile of \equ{velocity}.
We obtain
\begin{eqnarray}
{\dot M} &=& 4 \pi \rho_s r_s^2 c_s
= 4 \pi \rho_i r_i^2 v_i \\ \nonumber
&=& 4 \pi \rho_i r_s^2 c_s
\exp\left( {\phi(r_i) - \phi(r_s) \over c_s^2} - {1\over 2}\right) .
\end{eqnarray}
The wind solution is not strictly valid at $r\rightarrow 0$,
but since we chose $r_i$ such that $v\ll c_s$, we can relate
$\rho(r_i)$ to the gas density $\rho_0$ at $r=0$
via the static solution to eqs.~\ref{eq:cont} through \ref{eq:state},
\be
\rho_0 = \rho_i
\exp \left( {1\over c_s^2} [ \phi(r_i) - \phi(0)] \right)  .
\label{eq:static}
\ee
Thus, the mass loss rate is
\be
{\dot M} = 4 \pi \rho_0 r_s^2 c_s \exp
\left(-{1\over c_s^2} [ \phi(r_s) - \phi(0)] -{1\over 2}\right) .
\label{eq:mdot}
\ee

The total mass of gas within $r_s$, using \equ{static},
is approximately
\be
M_{gas}(r_s)
= 4 \pi \rho_0 \int_0^{r_s} r^2 \exp
\left( -{1\over c_s^2} [ \phi(r) - \phi(0)]\right) dr .
\label{eq:gas_mass}
\ee

The {\em evaporation time scale},
defined by $\tau_{evap} \equiv M_{gas}/\dot M$, can now be obtained
from
\equ{gas_mass}
and \equ{mdot}:
\be
\label{eq:evap}
\tau_{evap} =
{\sqrt{e} \over c_s}  \int_0^{r_s} \left( r\over r_s\right)^2
\exp\left( [ \phi(r_s) -\phi(r)]/c_s^2\right) dr  .
\ee
Note that it does not involve the values of the potential or the density
at the halo center.

To obtain the above result,
we implicitly assumed that the static solution, \equ{static},
is valid for the density out to $r_s$. This is a sensible approximation
as long as the bulk of the mass at $r<r_s$ has $v\ll c_s$.
Under this limit, where most of the mass is almost at rest,
the total mass estimate of \equ{gas_mass} is valid.
If we only have $v \leq c_s$, a large fraction of the
mass evaporates on an acoustic time scale --
the sound crossing time of the characteristic length $r_s$.
In this case we do not have a steady state and the system should be
treated in another limit,
as in \citet{barkana:99}.
Quantitatively, the requirement that our solution be valid is therefore
\be
\label{eq:validity}
\tau_{evap} \gg \tau_{acoustic} \equiv {r_s / c_s}.
\ee

In order to apply the above result, we need a potential profile and the
value of $c_s$ as input.
We then find the sonic point by \equ{sonic-def}, and then use
\equ{mdot} or \equ{evap}.

\section{Application to realistic haloes}
\label{sec:nfw}

\subsection{An NFW Profile}

We apply the above evaporation analysis to the typical profile
of galactic haloes, as seen in cosmological N-body simulations of the 
standard
$\Lambda$CDM cosmology. The dark-matter density is assumed to
follow the NFW profile \citep{navarro:97}:
\be
\rho_{dm}(r)
  =  {4 \rho_{c} \over {x}(1+x)^2} ,
\quad x\equiv {r \over \rc} ,
\ee
where $\rc$ is a characteristic inner radius\footnote{One may loosely
use the term ``core" radius even when it is actually a steeper ``cusp".}
and $\rho_c$ is the density at $\rc$.
The mass profile is obtained by an integral over $r$:
\be
M_{dm}(r) = 16 \pi \rho_c \rc^3 A(x), \quad
A(x) \equiv \ln (1+x) -{x \over 1+x} .
\ee
We may replace below the parameter $\rho_c$ by $M_{c}$,
the mass enclosed within $\rc$, which is given by
\be
\Mc = 8 \pi \rho_c \rc^3 (\ln 4 - 1).
\label{eq:mc}
\ee

Assuming that the gravity is dominated by the dark matter,
the gravitational force is
\be
f_{grav}(r) = - {G M_{dm}(r) \over r^2}
= - 16 G \pi \rc \rho_c {A(x) \over x^2},
\ee
and the potential is:
\be
\phi(r) =-\int_0^{r} f_{grav}(r)
= 16 G \pi \rho_c \rc^2 B(x)
\label{eq:phi_nfw}
\ee
where
\be
B(x) \equiv {x - \ln (1+x) \over x} .
\ee

The sonic point $r_s$ is the solution of \equ{sonic-def}, which takes 
the form:
\be
{A(x_s) \over x_s} = {(\ln 4-1) \over \psi}, \quad
\psi \equiv {G \Mc / \rc \over c_s^2} .
\label{eq:sonic_nfw}
\ee
The dimensionless parameter $\psi$
characterizes the general behavior of the gas in the given system. It
relates the gravitational potential at the core radius to the
specific thermal energy of the gas. In other words, it is the ratio 
squared
between the circular velocity at the core radius and the isothermal 
speed
of sound.
Depending on the value of $\psi$, the system could be in one of the
following three regimes.
A value of $\psi \gg 1$ implies that the gas is tightly bound
and the evaporation is negligible.
When $\psi > 1$ but $\psi$ is not too large, we expect a steady-state
evaporation by a thermal wind over many dynamical time scales.
On the other hand, when $\psi \lsim 1$, we expect no steady-state 
behavior
and the gas evaporates rapidly over a few dynamical time scales.
By solving \equ{sonic_nfw}, we obtain the dimensionless sonic radius
$x_s(\psi)$.

\begin{figure}
\hskip -0.5cm
\center{\epsfig{file=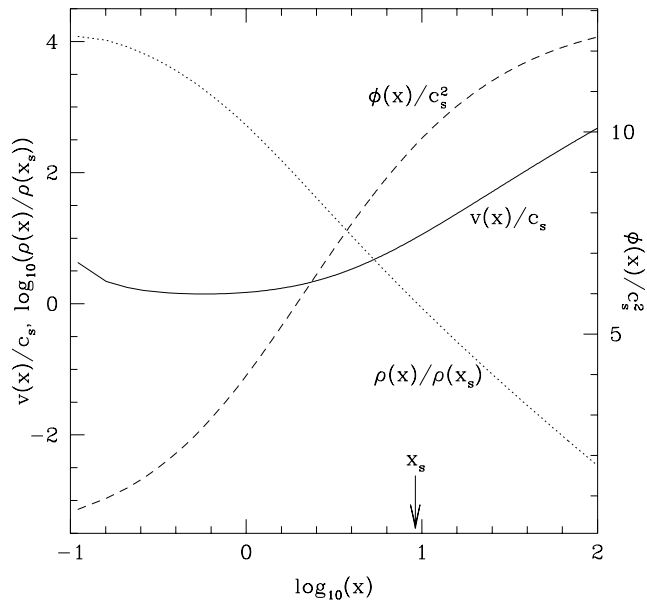,width=3.4in}}
\caption{ \small \sf
The solution for the evaporation of gas from an NFW potential,
with $\psi\equiv (G\Mc /\rc)/c_s^2 = 2.5$, in the regime where
the evaporation is through a steady wind.
The sonic point is at $x_s = 9.18$.
Shown are the NFW potential $\phi(x)$ in units of $c_s^2$,
the gas velocity in units of $c_s$,
and the gas density $\rho(x)$ in units of its value at the sonic point.
Note that since the sonic point is likely to be comparable to the virial
radius, our spherical solution may become invalid outside the sonic 
point.
This does not change the estimates of the mass loss, but it may change
the expected gas velocity at large radii.
}
\label{fig:sol1}
\end{figure}

Once $x_s$ is found, we can obtain the actual wind solution.
To do so, we take the NFW potential, \equ{phi_nfw},
and plug it in \equ{velocity},
which is an implicit equation for $v(r)/c_s$.
This equation can then be solved numerically. The solution for the
wind velocity profile $v(x)$,
together with the gas density profile $\rho(x)$ and the NFW potential,
are plotted in fig.~\ref{fig:sol1}
for the specific case of $\psi=2.5$, for which $x_s = 9.18$.
An interesting point to note is that
this sonic radius is comparable to the virial radius in a typical NFW 
halo,
which is on the order of $\Rv \sim 10 \rc$
\citep[see][]{bullock:01a}.
This implies that the wind solution at the sonic point may be perturbed
by deviations from spherical symmetry due to neighboring haloes.
This may have a noticeable effect on the velocity attained by the 
evaporated
gas, but since the sonic point is determined by conditions internal to 
it,
we expect no significant effect on the mass-loss rate itself.
This is because, by definition, material past the sonic point is moving
too fast to acoustically affect conditions up stream.

A useful dimensionless function of $\psi$
is the gas mass within $r_s$ in units of $(4 \pi /3) \rho_0 \rc^3$:
\begin{eqnarray}
f_1(\psi) &\equiv& { M_{gas} \over (4 \pi /3)\rho_0 \rc^3}\\ \nonumber
&=&  3  \int_0^{x_s(\psi)} x^2
\exp \left[ - {2\psi \over (\ln 4 -1)} B(x) \right] dx,
\end{eqnarray}
where $M_{gas}$ is taken from \equ{gas_mass} with the NFW potential
from \equ{phi_nfw}.
Note that $\psi$ enters both in the integrand and
through the upper limit of the integration $x_s(\psi)$.

Another useful dimensionless function of $\psi$ is
the mass-loss rate in units of $(4 \pi /3) \rho_0 \rc^3$
per the gas acoustic time scale $\rc/c_s$:
\begin{eqnarray}
f_2(\psi) &\equiv& {\dot{M} \over (4 \pi /3)\rho_0 \rc^3/ (\rc/c_s)}
\\ \nonumber
&=&  3   \exp \left[ -{1\over 2} - {2 \psi \over (\ln 4 -1)}
B[x_s(\psi)]  \right].
\end{eqnarray}

The evaporation time scale, for an $e$-fold mass loss, is therefore, as
in \equ{evap}, given by:
\be
\tau_{evap} = {M_{gas} \over \dot{M}}
= {\rc\over c_s} {f_1(\psi) \over f_2(\psi)}.
\label{eq:t_evap}
\ee

\begin{figure}
\hskip -0.5cm
\center{\epsfig{file=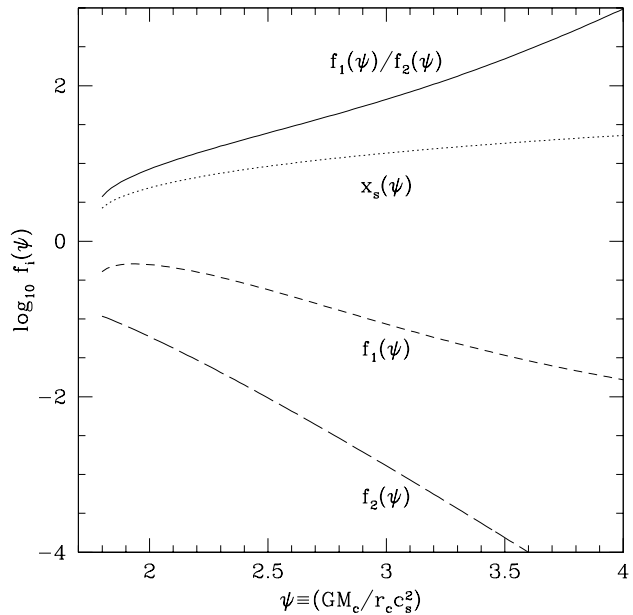,width=3.4in}}
\caption{ \small \sf
The dimensionless functions of $\psi$ for NFW haloes,
where $\psi$ characterizes the ratio of potential to thermal energy.
For $\psi < 1.78$ there is no steady state solution, and evaporation
takes place after a few sound-crossing times.
The function $x_s(\psi)$ is the radius of the sonic point in units of
the core radius.
The function $f_1(\psi)$ is the gas mass within the sonic point
in units of $4 \pi / 3 \rho_0 \rc^3$ where $\rho_0$
is the gas density at $r=0$.
The function $f_2(\psi)$ is the mass-loss rate in the above mass units
per sound crossing time $\rc/c_s$.
The ratio $f_1(\psi)/f_2(\psi)$ is the thermal evaporation time scale
in units of $\rc/c_s$.
}
\label{fig:funs}
\end{figure}

The solution of $x_s(\psi)$, together with the functions,
$f_1(\psi)$, $f_2(\psi)$, and $f_1(\psi)/f_2(\psi)$,
are depicted in fig.~\ref{fig:funs}, as a function of $\psi$.
The ratio $f_1(\psi)/f_2(\psi)$ gives the evaporation time scale in
units of the sound crossing time.
Take, for example, the case where $\psi = 3$.
Most of the gas is dynamically bound, as the potential energy dominates 
over
the thermal energy, but
the evaporation process could be important if enough time is available.
In this case,  $f_1(\psi)/f_2(\psi) \approx 50$, such that it takes 
about
50 sound crossing times of the core radius to evaporate the gas.

\subsection{Haloes of different mass and velocity}

Since we wish to compare the above results to actual cosmological 
haloes,
it is convenient to express the NFW profile parameters in terms of the
``virial" quantities:
the radius $\Rv$, the mass interior to it $\Mv$, and the circular
velocity at the virial radius $\Vv$.
Also, instead of using the more physically relevant acoustic time scale,
we should express the evaporation time in units of the Hubble time at
the corresponding cosmological epoch.

At redshift $z$, the virial mass is related to the virial radius through
the universal top-hat overdensity $\Delta$ at virialization:
\be
\Mv \equiv {4 \pi \over 3} \Delta(z) \rho_u \Rv^3,
\ee
where $\rho_u$ is the average universal density at $z$.
The corresponding virial velocity is defined by
\be
\Vv^2 \equiv {G \Mv \over \Rv }.
\ee

For the family of flat cosmologies ($\omm + \oml = 1$),
the virial overdensity can be approximated \citep{bryan:98} by:
\be
\Delta(z) \approx 
{\left( 18 \pi^2 + 82[\omm(z)-1] -39[\omm(z)-1]^2 \right) \over \omm(z)},
\ee
where $\omm(z)$ is the mass density parameter at redshift $z$.
It is related to today's values of the cosmological parameters by
\be
\omm(z) \approx {\om0 (1+z)^3 \over \om0 (1+z)^3 + 1-\om0}.
\label{eq:omm}
\ee
At the high redshifts of relevance to us here, $\omm$ is close to unity,
and $\Delta(z) \approx 180$.
For the standard $\Lambda$CDM cosmology, with $\om0=0.3$ and $\ol0=0.7$,
today's value is $\Delta(z=0) \simeq 340$.

If we define a factor of order unity by
\be
F_1 \equiv
{\om0 \over 0.3} {\Delta(z) \over 200} \left( {h \over 0.7} \right)^2,
\ee
where $h$ is the Hubble constant in units of $100 \kms {\rm Mpc}^{-1}$,
we can write the above virial relations (to an accuracy better than 
1\%) as
\be
M_8=0.342 F_1 (1+z)^3 R_{10}^3 ,
\ee
\be
V_{10}=0.383 F_1^{1/2} (1+z)^{3/2} R_{10} ,
\ee
\be
M_8=6.06 F_1^{-1/2} (1+z)^{-3/2} V_{10}^3 ,
\ee
where
$M_8 \equiv \Mv/10^8 M_\odot$,
$R_{10} \equiv \Rv/10 \kpc$,
and $V_{10} \equiv \Vv/10 \kms$.

To relate the virial variables to the core characteristics of the NFW
profile we use the concentration parameter, defined by
\be
C \equiv \Rv/\rc.
\ee
At a given $\Mv$ and $z$, we use here the average concentration 
parameter
as derived by \citet{bullock:01a} from an $N$-body simulation of the
$\Lambda$CDM cosmology.  It can be approximated by
\be
C(\Mv,z) \approx
20\, (1+z)^{-1}\left(\Mv \over 10^{11} M_{\odot}\right)^{-0.13}.
\ee

\begin{figure}
\hskip -0.5cm
\center{\epsfig{file=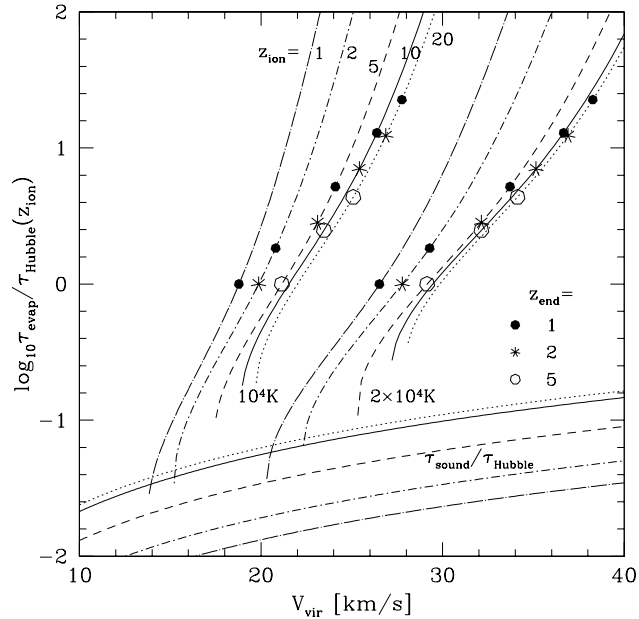,width=3.4in}}
\caption{ \small \sf
The evaporation time of haloes with different virial
velocities $\Vv$. The calculation is carried out at different redshifts
$\zion$ when the gas is assumed to be photo-ionized (as marked).
The evaporation time is
expressed in units of the Hubble time at that $\zion$.
The sets of curves on the left and on the right correspond to $T=10^4$K
and $2\times10^4$K respectively.
The velocity at which the curve reaches $\tau_{evap}/\tau_{Hubble}=1$
marks the critical virial velocity of a halo that loses an $e$-fold of
its gas in one Hubble time.
Also marked on each curve are the Hubble times at later redshifts 
$z_{end}$
at which the ionization terminates.
For example, a halo which is ionized to $10^4$K at $\zion=10$
(the solid curve in the left set),
and is subject to an effective ionizing flux until $\zend=1$ (the dot 
on the
curve), will decrease its gas mass by an $e$-fold if its virial velocity
is $\Vv=26.4\kms$.
If $T=2\times 10^4$K, a halo of $\Vv=36.8\kms$ will suffer a similar 
mass loss
by $z=1$.
The lower set of curves is the ``acoustic time scale" for the different
redshifts. Our wind calculation is valid as long as this time scale
is much smaller than $\tau_{evap}$. 
}
\label{fig:tauevap}
\end{figure}

Once we combine the above relations to the results obtained for the
thermal evaporation of the gas, we are in a position to
evaluate the evaporation process of typical cosmological haloes.
In particular, we can compare the evaporation time scale,
by which the halo has lost an $e$-fold of its gas mass,
to the Hubble-expansion time scale
at the onset of ionization, $\zion$.
This is depicted in fig.~\ref{fig:tauevap},
where we plot $\tau_{evap}/\tau_{Hubble}(\zion)$ for a set of initial 
and
terminal redshifts for the ionizing radiation.
 From the figure, it is apparent that for earlier values of $\zion$,
the thermal evaporation process can evaporate gas
from haloes with a somewhat higher virial velocity.
This is mainly because at a higher redshift a halo of a given $\Vv$
has a smaller $\psi$, roughly $\psi \prop (1+z)^{-0.8}$.
[It comes about because, for a given $\Vv$,
$\Rv\prop \Mv \prop (1+z)^{-3/2}$ and roughly $C \prop (1+z)^{-0.8}$,
such that $\rc\equiv \Rv/C \prop (1+z)^{-0.7}$.
On the other hand, $\Mv \prop (1+z)^{-3/2}$,
and $\Mc/\Mv$ is only a weak function of $C$. Together, with $c_s$ the 
same
at all redshifts, we get that
roughly $\psi \prop \Mc/\rc \prop (1+z)^{-0.8}$.]
We see in \Fig{funs} that $f_1/f_2$ is varying exponentially with 
$\psi$,
$f_1/f_2 \prop e^{b(1+z)^{-1/2}}$,
so we expect a high reduction in $f_1/f_2$ at a high $z$.
The multiplication by $\rc \prop (1+z)^{-0.7}$ in \equ{t_evap}
makes the decrease of $\tau_{evap}$ with $z$ even steeper.
The corresponding decrease in $\tau_{Hubble}$,
which at early times is $\prop (1+z)^{-3/2}$,
weakens the decrease of $\tau_{evap}/\tau_{Hubble}$ with $z$, but it
is not steep enough to compensate for the exponential
decrease in $\tau_{evap}$.
The net result is therefore an increase in the evaporation efficiency
as a function of $z$.
For example, if $t=10^4$K,
the critical velocity for $e$-fold loss in one Hubble time grows
from $\Vv=18.8\kms$ at $\zion=1$ to $\Vv=22.0\kms$ at $\zion=20$.
However, if the ionizing flux remains effective until, say, $z=1$,
the critical velocity for $\zion=20$ becomes $\Vv=27.7\kms$.

The terminal redshift, when the ionizing radiation is switched off,
is also a factor in
determining the critical halo virial velocity from which gas can 
evaporate.
For example, if $T=10^4$K and $\zion =10$,
haloes that lose an $e$-fold of their gas in either one Hubble time, by 
$z=5$,
$z=2$, or $z=1$, are of $\Vv\approx 21.5$, $22.3$, $25.5$, and 
$26.4\kms$
respectively.

On the other hand, the temperature at which the gas can be kept
heated is more important at determining the evaporation losses.
This is because it determines the speed of sound.
Since $c_s \propto \sqrt{T}$, all typical velocities in the system scale
in the same manner. Namely, by increasing $T$ from $10^4$K to $2 \times 
10^4$K,
the critical velocity for $e$-fold evaporation becomes larger
by a factor of $\sim \sqrt{2}$.
For example, if $\zion=10$, the critical velocity for an $e$-fold loss 
in
a Hubble time grows from $\Vv=21.5 \kms$ for $T=10^4$K to $\Vv=29.3$ for
$T=2\times 10^4$K.

The intersections between the 
acoustic curves and the evaporation curves plotted in fig.~\ref{fig:tauevap},
correspond to the lowest velocities for which
there is a consistent steady state thermal evaporation, and $e$-fold 
mass loss over a dynamical time. These velocities are somewhat larger than the 
velocities corresponding to instantaneous mass loss found by \citet{barkana:99}
using binding energy arguments. 

\begin{figure}
\hskip -0.5cm
\center{\epsfig{file=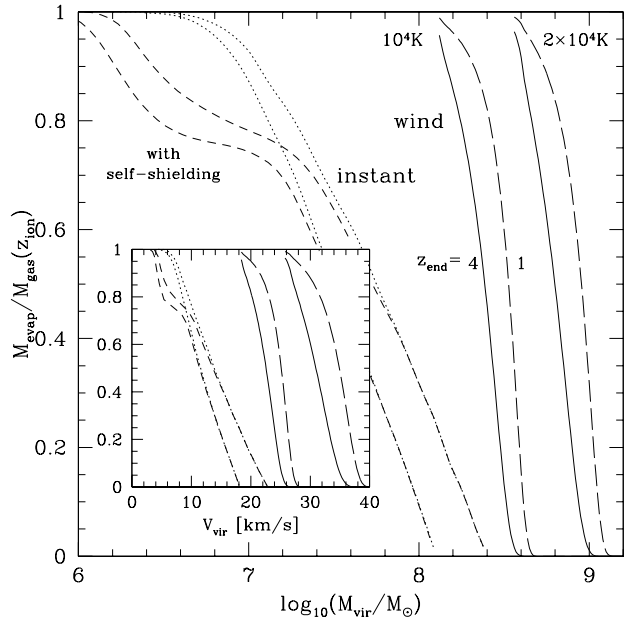,width=3.4in}}
\caption{ \small \sf
Fraction of gas mass evaporated from haloes due to re-ionization at 
$\zion=8$
as a function of halo mass or virial velocity (inset).
The two pairs of curves on the right are the results of evaporation by 
thermal
wind continuing until $\zend=4$ and $1$,
for $T=10^4$ and $2\times 10^4K$ as marked.
The two pairs of curves on the left are the gas fraction
evaporated instantaneously according to \citet{barkana:99}.
In each pair, the curves are for $T=10^4$ (left) and $2\times 10^4K$ 
(right).
The pair of curves on the very left, showing less evaporation at small 
masses,
represents the inclusion of self shielding,
which enables the existence of a dense core inaccessible to the
effects of the external ionizing radiation.
We find that the continuous evaporation by wind increases the critical 
mass
for evaporation by an order of magnitude, and it sharpens the transition
between efficient and inefficient evaporation.
}
\label{fig:MevapMvir}
\end{figure}

The fraction of gas not evaporated decreases exponentially with time,
such that over a given time $\Delta t$, this fraction is
$\exp(-\Delta t/\tau_{evap})$. Thus, using the evaporation time scale,
we can estimate the amount of gas that will evaporate over a Hubble 
time.
This is described in fig.~\ref{fig:MevapMvir}, which plots the fraction
of evaporated gas as a function of the virial mass of the halo,
or the virial velocity in the inset.
Also plotted is the fraction of gas which becomes unbound
and ``instantly" evaporates on a dynamical time scale once
ionization is switched on, as calculated by \citet{barkana:99}.

 From the figure, it is evident that by letting the slow process of
thermal evaporation take place,
a given relative mass loss is obtained in
haloes which are typically 10 times more massive.
For example, only haloes of $\lsim 2\times 10^7 M_{\odot}$ ($\lsim 10 
\kms$)
lose 70\% of their gas mass from the instantaneous evaporation,
while the thermal wind ensures that haloes with
$\lsim 2\times 10^8 M_\odot$ ($\lsim 22\kms$) lose 70\% of their gas 
mass, if $T=10^4 K$. For $T=2\times 10^4 K$, the equivalent mass 
increases by more than a factor of two.
In the case of
haloes which suffered higher mass loss fractions,
we find that the inclusion of thermal evaporation can increase the
equivalent cutoff virial mass by almost 100-fold.
This is because the thermal wind can even evaporate the dense cores
which otherwise are shielded from the ionizing radiation
and therefore do not evaporate dynamically.

We also see in the figure that the process has a sharp cutoff,
that is, either a halo can or it cannot retain its gas.
Only a very narrow range of halo masses partially retain their
original gas content.
Also apparent is that the critical halo mass is not sensitive to the
epoch when the ionization becomes ineffective. We learned in 
\Fig{tauevap}
that it is also not too sensitive to the epoch when the ionization 
starts,
but it is quite sensitive to the temperature at the ionized state.

\subsection{Mass budget}

\begin{figure}
\vskip 0.95cm
\center{\epsfig{file=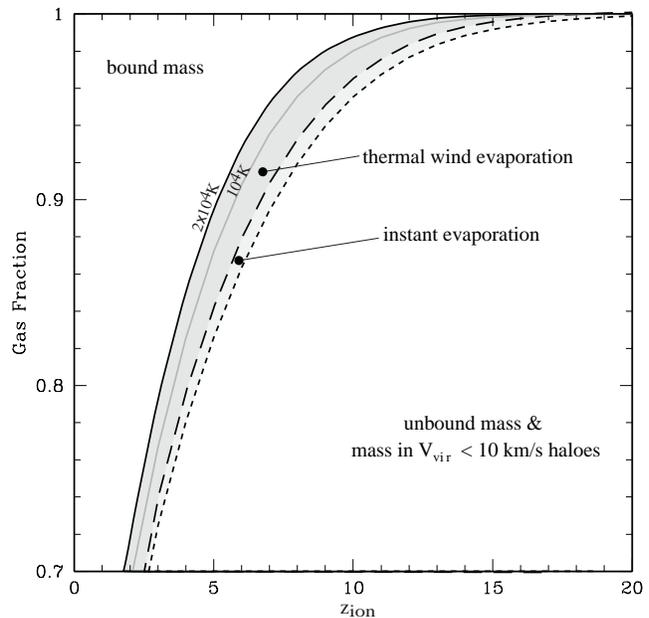,width=3.3in}}
\vskip 0.2cm
\caption{ \small \sf
``Mass budget" at the onset of reionization $\zion$.
The area below the shaded area refers to gas
which is either originally associated with haloes of $\Vv<10\kms$ or
not associated with any virialized halo.
This gas is not involved in star-forming galaxies.
The gas above the shaded area is bound to massive haloes and
it will not photo-evaporate by $\zend=2$.
The shaded area refers to the gas that will evaporate from $\Vv>10\kms$ 
haloes.
The top two curves refer to continuous evaporation by $\zend=2$
for the two possible temperatures as marked.
The shaded area between the dashed lines refers to gas that escapes on a
dynamical time.
One may assume that
the gas in the shaded strip is recycled in dwarf galaxies
and returned to the IGM.
}
\label{fig:budget}
\end{figure}

Next, we wish to evaluate the total fraction of gas evaporated from 
haloes
during the ionization period.
We consider the distribution of halo masses at $\zion$
and integrate over the fraction of gas evaporated from each of these 
halos,
either instantaneously or under the assumption that the ionization 
remains
effective until $\zend=2$.
The results are shown in \Fig{budget}.

The distribution of halo masses is estimated using the Sheth-Tormen
modification \citep{sheth:99} of the Press-Schechter approximation.
Assuming a cosmological model and Gaussian initial fluctuations
with a given linear power spectrum,
the fraction of mass in haloes of virial mass larger than $M$
is estimated by
\be
F(>M,z) = 0.4 \left( 1+{0.4\over \nu^{0.4}} \right)
{\rm erfc}\left({0.85 \nu \over \sqrt{2}}\right),
\ee
where
$\nu \equiv \delta_c/[D(z) \sigma(M)]$,
with $\delta_c=1.69$, $D(z)$ the linear fluctuation growth rate,
and $\sigma(M)$ the $rms$ amplitude of linear density fluctuations
today top-hat smoothed over mass $M$.
We assume the standard flat $\Lambda$CDM cosmology with $\om0=0.3$
and $\ol0=0.7$.
Following \citet{carroll:92}, we approximate
\be
D(z) \approx {\om0^{4/7}+(3/2)\om0 \over \om0}
             {\omm(z) \over [\omm(z)^{4/7}+(3/2)\omm(z)] (1+z)} ,
\ee
with $\omm(z)$ from \equ{omm}.
The $rms$ fluctuation $\sigma(M)$ is derived in the standard way
using a top-hat window and the CDM power spectrum of \citet{bardeen:86} 
with
baryonic density $\omb=0.044$,
Hubble constant $h=0.7$,
power index $n=0.95$ and
the amplitude of the power spectrum normalized by $\sigma_8=0.85$
(motivated by the best fits from the CMB observations by the MAP 
satellite).

In \Fig{budget}, for every value of $\zion$, the gas is divided into
three different populations.  Below the shaded area is the gas
that either has evaporated from mini-haloes of $\Vv<10\kms$
(where the inefficient cooling does not allow any stars to form)
or ``field gas" that has not been in any virialized halo to begin with.
The shaded area refers to the fraction of the gas
at $\zion$ that will be photo-evaporated by $\zend=2$ from
haloes of $\Vv>10\kms$.
The region above the shaded area is the rest of the gas, bound to 
massive
haloes and not to be photo-evaporated.
The evaporated fraction is further divided into gas that escapes on
a dynamical time scale at $\zion$ versus gas that flows out continuously
in a thermal wind until $\zend=2$.
The results for the two possible gas temperatures are shown.

We find that quite independently of $\zion$,
the amount of gas which collapsed into virialized objects and later 
evaporated
is about 6\% if the temperature of the gas was kept at $\sim 2\times 
10^4$K,
or about 4\% if $T$ was kept at $\sim 10^4$K.
The instant evaporation involved 1 to 1.5\% of the gas.
In other words, the photo-evaporation process recycled gas in dwarf 
galaxies,
and it can therefore have implications on the metallicity of the IGM.

\subsection{The assumption of isothermality}
\label{sec:isothermality}

The results thus far described should be considered cautiously until
we verify the basic assumption behind the evaporation process ---
that the gas is kept isothermal during the evaporation process.
This implicitly assumes that the typical time scale for gas evaporation
is longer than the time it takes the expanding
gas to heat and establish equilibrium with the radiation field.
If this assumption fails, the gas will expand adiabatically
instead and cool rapidly with distance.  Without the extra energy input
from the radiative heating, the net result will be significantly less
mass loss from given haloes.

Under the assumption that the ionizing radiation is strong enough to
keep the gas hot and in equilibrium between heating and cooling,
any additional assumption regarding the heating rate is equivalent to an
assumption regarding the cooling rate. Thus, we can test instead the 
condition
that the cooling time is short enough, as compared with the time it 
takes
the gas to outflow from the potential well and reach the sonic point.
This is typically the sound crossing time at the virial radius.

 From the results of \citet{sutherland:93,katz:96},
we find that for temperatures in the range $10^4$K and $10^5$K
and for {\em primordial composition} the cooling rate
is $\Lambda = 5 \times 10^{-23} \Lambda_0  
n_H^2\mathrm{erg~s^{-1}cm^3}$,
where $n_H$ is the number density of the gas and
$\Lambda_0 \approx 1$ to within a factor of two
(If the gas is contaminated by metals, the cooling rate could be even 
faster).
The characteristic cooling time can then be defined by
$\tau_{cool} \equiv (3/2) c_s^2 \rho / \Lambda$,
where $c_s$ is the {\em isothermal} speed of sound, corresponding to a
temperature $T$.
The sound crossing time at the virial radius defines the
dynamical time, $\tau_{dyn}\equiv R_v/c_s$.
We write the gas density as
$\rho = \Delta_b \Omega_b \rho_{crit,0} (1+z)^3$,
where $\Delta_b$ is the overdensity of baryons at the virial radius,
and $\rho_{crit,0}$ is the critical density of the universe today.
Adding these expressions together results with:
\be
{\tau_{dyn} \over \tau_{cool}} \approx 1.5 F_2
\left({\Vv \over 10\kms} \right)
\left({\Delta_b \over 10} \right)
(1+z)^{3/2} ,
\ee
where $F_2$ is a factor of order unity,
\be
F_2 \equiv
\left({\omm \over 0.3} {\Delta(z) \over 200} \right)^{-1/2}
\left({h \over 0.7 }\right)
\left({\omb \over 0.05} \right)
\left({T \over 10^4 {\rm K}} \right)^{-3/2}
\Lambda_0.
\ee
In the range of interest we have $\Vv \sim 20\kms$ and $z > 1-2$,
so $\Delta_b >10$ would guarantee efficient cooling.
We can evaluate $\Delta_b$ in the case where no wind is present
by inspecting Fig.~2 of \citet{barkana:99}, which shows that
$\Delta_b$ is typically $\gsim 10$.
The presence of a wind can only increase the density
at the sonic point, because in this case the density
falls off at large distances as a power law, while in the no-wind
solution it falls off exponentially.
We therefore expect $\Delta_b > 10$, and conclude that the
condition for isothermality should be valid.
Before $z \sim 4$, we expect the cooling time to be shorter than the
dynamical time by more than an order of magnitude.

\section{Discussion}
\label{sec:conc}

We analyzed the continuous evaporation from dwarf galaxies by thermal 
winds during the period when an ionizing flux was effective.
We found that the critical virial velocity for significant evaporation 
is significantly higher than computed before assuming an instantaneous 
mass loss.
For example, if the ionization starts at $\zion=10$ and is maintained until
$\zend=2$, a mass loss of one $e$-fold occurs in haloes of $\Vev \simeq 24$ to
$34 \kms$, depending on whether 
Hydrogen or also Helium are reionized respectively.
One $e$-fold of gas is lost within the first Hubble time at $z=10$ for 
haloes of $\Vev \simeq 20$ to $29\kms$ respectively.

Before we continue with the discussion of possible implications, we 
should mention the main caveats in our analysis.
First, we assumed spherical symmetry for the system of dark halo and gas.
This is generally not strictly valid, because the galaxies are typically
triaxial configurations which may also involve significunt substructure.
The fact that the equi-potential surfaces tend always to be rounder 
than the
equi-density surfaces helps making the spherical approximation
sensible for our analysis. Nevertheless, this is a limitation on the
accuracy of our results.

The other caveat, as discussed before,
is the fact that the sonic radius obtained
is typically comparable to the halo virial radius.
On scales larger than the virial radius, one may expect the potential
wells to deviate from the NFW profile and be perturbed by neighboring
structure. In fact, the potential well need not be spherically 
symmetric there,
nor should it even continue to increase at very large distances.
This implies that one should not take too seriously the wind solution at
$r > r_v \sim r_s$. For example, the formal solution shown in \Fig{sol1}
indicates that the wind continues to accelerate past the
sonic point to more than twice the sound speed at large radii.
This will not necessarily be the case. Nevertheless, once the wind
turns super-sonic, it obviously cannot affect conditions present 
upstream.
This implies that as long as the sonic point or regions internal to it 
are not affected by nearby haloes, conditions at the sonic point which
determine the mass loss rate, will not change.

As long as the IGM is ionized, accreting gas is not expected to quench the 
steady thermal winds. This is because the Jeans scale corresponds
to haloes of virial velocity $\sim 30\kms$ or larger 
\citep[e.g.][]{barkana:01}, implying that no significant gas infall is 
expected into haloes that support a steady wind.

The upper critical velocity for evaporation 
is thus
comparable to the
Jeans scale, and significantly larger than the cooling barrier at
$\sim 10\kms$, below which it is believed that luminous galaxies cannot 
form.  
This implies that 
every halo in the crude range $10-30\kms$ is likely to be
affected by the photo-evaporation process. The mass fraction in such 
haloes is on the order of 5\% at any given time after $z\sim 10$.
Some of them eventually merge into bigger haloes, but some end up as
dwarf haloes today.
Among these unmerged haloes, those that managed to form some stars
prior to $\zion$
possibly make low-surface-brightness dwarf galaxies which are
dominated by an old population of stars and with no original gas 
retained.
Young stars can form from newly accreted gas after the ionization stops.
These objects may resemble the dwarf spheroidals of the Local Group.
The other haloes in the range $10-30\kms$, those that are dynamically 
younger
and did not form stars prior to $\zion$, were doomed to never form 
stars,
at least until $\zend$.
They are likely to end up as completely dark haloes.
These are therefore not limited to the mini haloes below the cooling 
barrier
of $\sim 10\kms$, but they rather extend up to $\sim 20-30\kms$.
This may help explain the apparent discrepancy between the predicted
mass function and the observed luminosity function at the faint end.
While certain variants of the currently accepted cosmological model
may predict a number density for subhalo satellites in big haloes
that matches the observed population of Local-Group dwarf galaxies
in the range $V >20-30\kms$, it is much harder for such models to 
recover
the small number at $V<20\kms$ \citep{zentner:03}.
This can be explained by the very efficient photo-evaporation of gas
taking place in haloes below $\Vev \sim 20\kms$, as found by our wind
analysis.

Given our results, we may attempt to address additional aspects of the
possible scenario for dwarf spheroidal galaxies.
They are constrained to form in haloes which lie in the crude range
$10\leq V \leq 30\kms$.
The gas in haloes of $V < 10 \kms$ cannot cool to form stars at any 
early
epoch because the cooling rate has a sharp drop below $T \simeq 10^4$K;
any molecular hydrogen that is necessary for cooling at $T<10^4$K is
likely to be dissociated even before reionization.
Efficient cooling by Hydrogen recombination at $T\gsim 10^4$K can
lead to an early burst of stars in haloes of $V \gsim 10\kms$.
The associated supernovae blow out part of the gas, as they do
with a gradually decreasing efficiency
in all haloes of $V<100\kms$ \citep{dekel:86,dekel:03}.
In $V < 30\kms$ haloes, the rest of the gas photo-evaporates in
a few Hubble times, as computed above.
No further gas can fall in as long as the ionizing flux is effective,
which may last until $z\sim 1-2$.
This leads to a gas-poor, old-population stellar system.
Any recent star formation in these dwarf spheroidals must be due to
gas falling in after the ionizing flux became ineffective.

An inspection of the global properties of dwarf galaxies indicates that
the dwarfs in the range $10-30\kms$ span
more than two orders of magnitude in luminosity, or stellar mass $\Ms$
\citep{dekel:03}. They show no significant correlation between velocity
and luminosity, contrary to the irregular, disky dwarfs that are 
typically
larger than $\sim 30\kms$ and tend to lie on an extension of the tight
Tully-Fisher relation defined by the brighter galaxies.
This observed property is consistent with the scenario sketched above
for the dwarf spheroidals.
The spread in $\Ms$ within this family
represents variations in quantities other than halo velocity,
such as the proximity of the earlier star formation epoch to $\zion$.
The spread in $\Ms$ for a given halo $V$ becomes
much smaller for the larger dwarf irregulars and bright galaxies
because haloes above $\Vev$ retain a larger fraction of their gas
and being above the Jeans scale they allow further accretion of more 
gas.
Star formation is thus allowed to occur in these objects
also after $\zion$, such that a large
fraction of the remaining gas can turn into stars. This brings the 
value of
$\Ms$ close to the value predicted by the energy requirement
for gas removal by supernova, which is a strong function of the halo
virial velocity: $\Ms/M \prop V^2$ \citep{dekel:03}.

The above scenario may also explain why the dwarfs at the low end,
$10-30\kms$, are typically spheroidals with low spin, while the larger
dwarfs tend to be disks.
According to the above scenario, the stars in the smaller galaxies are 
made
from the gas that cooled first, namely the gas residing in the inner 
halo.
This gas is typically of specific angular momentum significantly
lower than the average for that halo \citep{bullock:01b,maller:02}.
In fact, the scenario predicts a correlation between $\Ms$ and stellar
spin parameter because the low-$\Ms$ dwarfs typically
form later, closer to $\zion$. The smaller fraction of gas involved
in the fainter dwarfs must have originated from inner halo regions
which host the low-end tail of the specific angular-momentum 
distribution.
There is a hint for such a correlation in the Local Group dwarf 
spheroidals
\citep{dekel:03}.

The photo-evaporation process discussed in this paper may involve a
non-negligible fraction of the gas in the universe.
If it evaporated from dwarfs after
they formed stars, it could carry metals with it and enrich the IGM.
This could provide a lower limit to the actual enrichment, which is 
probably
dominated by supernova blowout.
We estimated for the standard $\Lambda$CDM cosmology
that at any given time about 5\% of the mass is in haloes
that can allow significant evaporation.
Integrated over time, we find that by $z \sim 1$
about 40\% of the mass is in haloes of $V>10\kms$. A significant 
fraction
of this mass must have been in haloes in the evaporating range for more
than a dynamical time sometime during the merger history. This
implies that up to 40\% of the gas has been evaporated from dwarf
galaxies, and could carry metals.

\section*{Acknowledgments}
This research has been supported
by the Israel Science Foundation grants 201/02 
and 213/02, by the German-Israel Science Foundation grant I-629-62.14/1999,
and by NASA ATP grant NAG5-8218.


\label{lastpage}

\begin{thebibliography}{}
\def\refe{\bibitem}

\refe[Babul \& Rees(1992)]{babul:92}
Babul A., Rees M.J., 1992, MNRAS, 255, 346
\refe[Bardeen \etal(1986)]{bardeen:86}
Bardeen J.M., Bond J.R., Kaiser N., Szalay A.S., 1986, ApJ, 304, 15
\refe[Barkana \& Loeb(1999)]{barkana:99}
Barkana R., Loeb A., 1999, ApJ, 523, 54
\refe[Barkana \& Loeb(2001)]{barkana:01}
Barkana R., Loeb A., 2001, PhR, 349, 125
\refe[Brandt(1970)]{brandt:70}
Brandt J.C., 1970, Introduction to the solar wind,
Series of Books in Astronomy and Astrophysics, Freeman, San Francisco
\refe[Bryan \& Norman (1998)]{bryan:98}
Bryan G., Norman M., 1998, ApJ, 495, 80
\refe[Bullock \etal (2001a)]{bullock:01a}
Bullock J.S., Kolatt T.S., Sigad Y., Somerville R.S., Kravtsov A.V.,
Klypin A.A., Primack J.R., Dekel A., 2001a, MNRAS, 321, 559
\refe[Bullock \etal (2001b)]{bullock:01b}
Bullock J.S., Dekel A., Kolatt T.S., Kravtsov A.V., Klypin A.A.,
Porciani, C., Primack J.R., 2001b, ApJ, 555, 240
\refe[Carroll, Press \& Turner(1992)]{carroll:92}
Carroll S.M., Press W.H., Turner E.L., 1992, ARA\&A, 30, 499
\refe[Dekel \& Silk(1986)]{dekel:86}
Dekel A., Silk J., 1986, ApJ, 303, 39
\refe[Dekel \& Woo(2003)]{dekel:03}
Dekel A., Woo J., 2003, submitted (astro-ph/0210454)
\refe[Haiman, Rees \& Loeb(1996)]{haiman:96}
Haiman Z., Rees M.J., Loeb A., 1996, ApJ, 467, 522
\refe[Katz, Weinberg \& Hernquist(1996)]{katz:96}
Katz N., Weinberg D.H., Hernquist L., 1996, ApJS, 105, 19
\refe[Kepner, Babul \& Spergel(1997)]{kepner:97}
Kepner J.V, Babul A., Spergel D.N., 1997, ApJ, 487, 61
\refe[Lamers \& Cassinelli(1999)]{lamers:99}
Lamers H.J.G.L.M., Cassinelli,  J.P., 1999,
Introduction to stellar winds, Cambridge University Press
\refe[Loeb \& Barkana(2001)]{loeb:01}
Loeb A., Barkana R., 2001, ARA\&A, 39, 19
\refe[Maller \& Dekel(2002)]{maller:02}
Maller A.H., Dekel A., 2002, MNRAS, 335, 487
\refe[Mateo(1998)]{mateo:98}
Mateo M., 1998, ARA\&A, 36, 435
\refe[Miralda-Escude \& Rees(1998)]{miralda:98}
Miralda-Escude J., Rees, M.J., 1998, ApJ, 497, 21
\refe[Miralda-Escude(2000)]{miralda:00}
Miralda-EscudŽ  J., 2000, ApJ, 528, L1
\refe[Navarro \etal (1997)]{navarro:97}
Navarro J.F., Frenk C.S., White S.D.M., 1996, ApJ, 462, 563
\refe[Parker(1960)]{parker:60}
Parker E.N., 1960, ApJ, 132, 821
\refe[Press \& Schechter(1974)]{press:74}
Press W.H., Schechter P., 1974, ApJ, 187, 425
\refe[Sheth \& Tormen(1999)]{sheth:99}
Sheth R.K., Tornem G., 1999, MNRAS, 308, 119
\refe[Sutherland \& Dopita(1993)]{sutherland:93}
Sutherland R., Dopita M., 1993, ApJS, 88, 253    
\refe[Woo \& Dekel(2003)]{woo:03}
Woo J., Dekel A., 2003, in preparation
\refe[Zetner \& Bullock(2003)]{zentner:03}                         
Zentner A.R., Bullock J.S., 2003, ApJ, submitted (astro-ph/0304292)

\end{thebibliography}
\end{document}